\documentstyle[prl,epsfig,aps]{revtex}

\begin{document}
\draft

\title{Study of the isotropic contribution to the analysis of 
photoelectron diffraction experiments at the ALOISA beamline}

\author{F. Bruno$^{a}$, L. Floreano$^{a,*}$, A. Verdini$^{a}$,
D. Cvetko$^{a,b}$, R. Gotter$^{a}$, A. Morgante$^{a,c}$.
}

\address{
$^{a}$Laboratorio TASC dell'Istituto Nazionale per la Fisica della 
Materia, S.S.14 Km 163.5, Basovizza, 34012 Trieste, Italy\\
$^{b}$J. Stefan Institute, 
University of Ljubljana, Slovenia, and Sincrotrone Trieste, 
Italy.\\
$^{c}$ Dipartimento di Fisica dell'Universit\`a di Trieste, 
Italy. \\
}
\author{M. Canepa, S.Terreni}
\address{INFM and Dipartimento di Fisica dell'Universit\`a di Genova,
Italy.}
\narrowtext
\onecolumn

\maketitle
\begin{abstract}

The angular distribution of the intensity in photoemission experiments 
is affected by electron diffraction patterns and by a smoothly varying 
contribution originated by both intrumental details and physical 
properties of the samples.
The peculiar variety of  scattering configurations available at the 
ALOISA beamline experimental station in Trieste  stimulated the 
developement of an analytical description for the smooth angular 
dependence sustaining the diffraction features.
We present here the basic formulae and their application to 
experimental data taken on the Fe/Cu$_{3}$Au(001) 
system in order to highlight the role of the various parameters 
included in the distribution function. A specific model for the 
surface illumination has been developed as well as the overlayer 
thickness and surface roughness have been considered.

 \\

$^{*}*$ Corresponding author: Fax: +39-04-226767,
\textit{E-mail address:} floreano@tasc.infm.it

\end{abstract}

\onecolumn
 

\widetext

\section{Introduction}
The study of the angular distribution of the Auger- and photo--electrons 
at medium kinetic energies (a few hundreds of eV) is a well 
established method to determine the atomic structure in a local 
environment surrounding the emitting atoms close to the surface of 
ordered materials.
The lattice geometry is usually retrieved exploiting the observation of 
intensity maxima along the direction of close packed atom rows. By taking into 
account the electron diffraction (ED) features, one can also reconstruct the 
atomic distances between the emitter and its closest neighbors.

In general, the distribution of the intensity in the ED patterns is originated by 
two contributions: one anisotropic component $\chi$, which 
is determined by the geometry of the local atomic structure, and one, 
slowly varying, isotropic component $ISO$ which depends from 
both instrumental factors (such as sample illumination and detector 
angular resolution) and material dependent factors (such 
as atomic differential cross section, film thickness/escape depth, 
surface morphology/roughness). 

The origin of the various contributions to the $ISO$ component has been 
identified since many 
years \cite{cooper,fadley74,litt}. Some aspects relevant to photoelectron 
diffraction have been recently reviewed \cite{woodruff}, in particular 
regarding the role of dipolar emission in surface science experiments.
Nonetheless in this work we present 
original developement of the ED analysis, which arises from the 
evolution of instrumental performance, in terms of analyzers 
positioning and angular resolution, as well as collimation and size of 
X-ray beams in third generation synchrotron sources.
The analytical treatement of the instrumental factors 
is presented in detail for  the end station of the
ALOISA beamline (Trieste 
Synchrotron), where a wide variety of scattering geometries is
available for  ED experiments.

In the following, we present a functional form of the $ISO$ 
component using physically meaningful parameters, 
that can be fitted at once with the $\chi$ 
component to rigorously deal with the ED analysis. 
This functional form has 
been coupled to multiple scattering calculations of the anisotropic 
structure-dependent $\chi$ component (we made use of the MSCD code by Chen and 
Van Hove \cite{MSCD}). This fitting procedure was applied to the 
structural study of thin Fe films grown on the Cu$_{3}$Au(001) 
surface, where the Fe structure changes from an fcc-like configuration 
to a bcc-like one as a function of the film 
thickness.\cite{kirschner,wuttig} 
For this system, Auger- and photo-electron ED data where taken in combination with 
in-plane X-ray diffraction.\cite{bruno} 
The latter technique yielded the lateral 
lattice parameter of the growing film with utmost precision and its 
value was used as input in the fitting of the ED patterns, the 
vertical lattice spacing thus being the only structural parameter to be 
determined. In the next section, we have compared our model of the $ISO$ 
function with the 
ED data taken for a few Fe films to enlight the weight of a proper choice 
of the $ISO$ functional form in the determination of the $\chi$ component. 

\section{Experimental setup}

The experimental chamber of the ALOISA beamline hosts several 
electron and photon detectors in UHV for the study of 
surface photoemission, X-ray and photolectron diffraction. This 
experimental setup exploits 
the very wide photon energy range given by the ALOISA grating-crystal 
monochromator, which spans from the dominion of 
X-ray photoemission spectroscopy (hundreds of eV) to that of X-ray 
diffraction (thousands of eV) \cite{floreano}.
The three angular degrees of freedom  
given by the manipulator (a modified CTPO manipulator by Vacuum 
Generators) allow one to freely orient the sample surface with 
respect to the photon beam direction and the beam linear polarization. 
At the same time, the combined rotation of the 
frame (rotation $B$) hosting in UHV the electron analyzers, and of the whole 
experimental chamber (rotation $C$) allows one 
to explore a wide portion of the hemisphere above the surface 
for any sample orientation.
(see Fig.~\ref{geometry} for a 
sketch of the scattering geometries accessible by the detectors).
The electron analyzers are of the hemispherical type with a 33~mm mean 
radius and an optical lens system yielding an acceptance angle of 
$\sim 1^{\circ}$~ (FWHM) and a field of view of $\sim 1 \times 
4$~mm$^{2}$ \cite{gotter}. The transverse size (FWHM) of the photon beam at 
the sample position is $\sim 20$~$\mu$m vertically and $\sim 
150$~$\mu$m horizontally, with a slight dependence on the photon 
energy. The photon beam is linearly polarized, with 
the electric field vector $\vec{E}$ in the horizontal plane 
\cite{aloweb}.

ED polar scans 
are usually collected by rotating one electron analyzer in the 
scattering plane from the surface normal to the horizon (polar angle 
$\theta$), even though different geometries can be suitable to 
 specific experimental requirements, such as photoelectron holography 
 \cite{holo} or near node photoelectron diffraction \cite{verdini}. 
 As far as the data reported in this paper are concerned, 
 the scans have been taken in transverse--magnetic 
polarization with the substrate [001] direction oriented along the 
photon beam direction for a few selected grazing angles $\alpha$~ 
on the surface. In the present case, the ED patterns have been collected 
for the Fe L$_{23}$M$_{23}$M$_{45}$ Auger line at a kinetic energy 
of 698~eV (we used a photon energy of 900~eV). In one case, we 
measured the ED pattern from a policrystalline Fe sample for the 3p 
core level at a kinetic energy of 803~eV. 
The signal was always taken at the maximum of the 
corresponding spectral line and at two suitably chosen energies aside the 
peak, in order to allow an effective subtraction of the background of 
secondary electrons.
The films were prepared by in situ evaporation of Fe on a 
Cu$_{3}$Au(001) surface and the thickness was measured in situ by 
X-ray reflectivity (further detail about the sample preparation can 
be found elsewhere \cite{bruno}). After an initial pseudomorphic 
stage, the Fe films assume a tetragonally distorted bcc structure,
displaying a (001) surface orientation, 
but rotated by 45$^{\circ}$ with respect to the substrate. 
At the largest thickness considered here (36~\AA), the Fe film has almost 
recovered its bcc bulk structure, 
with a lateral lattice spacing $a = 2.830 \pm 0.005$~\AA, as measured by in-plane 
X-ray diffraction \cite{bruno}.

\section{Description of the {\it ISO} function}

\paragraph{Angles definition}

\begin{figure}[tbp]
\begin{center}
    \epsfig{file=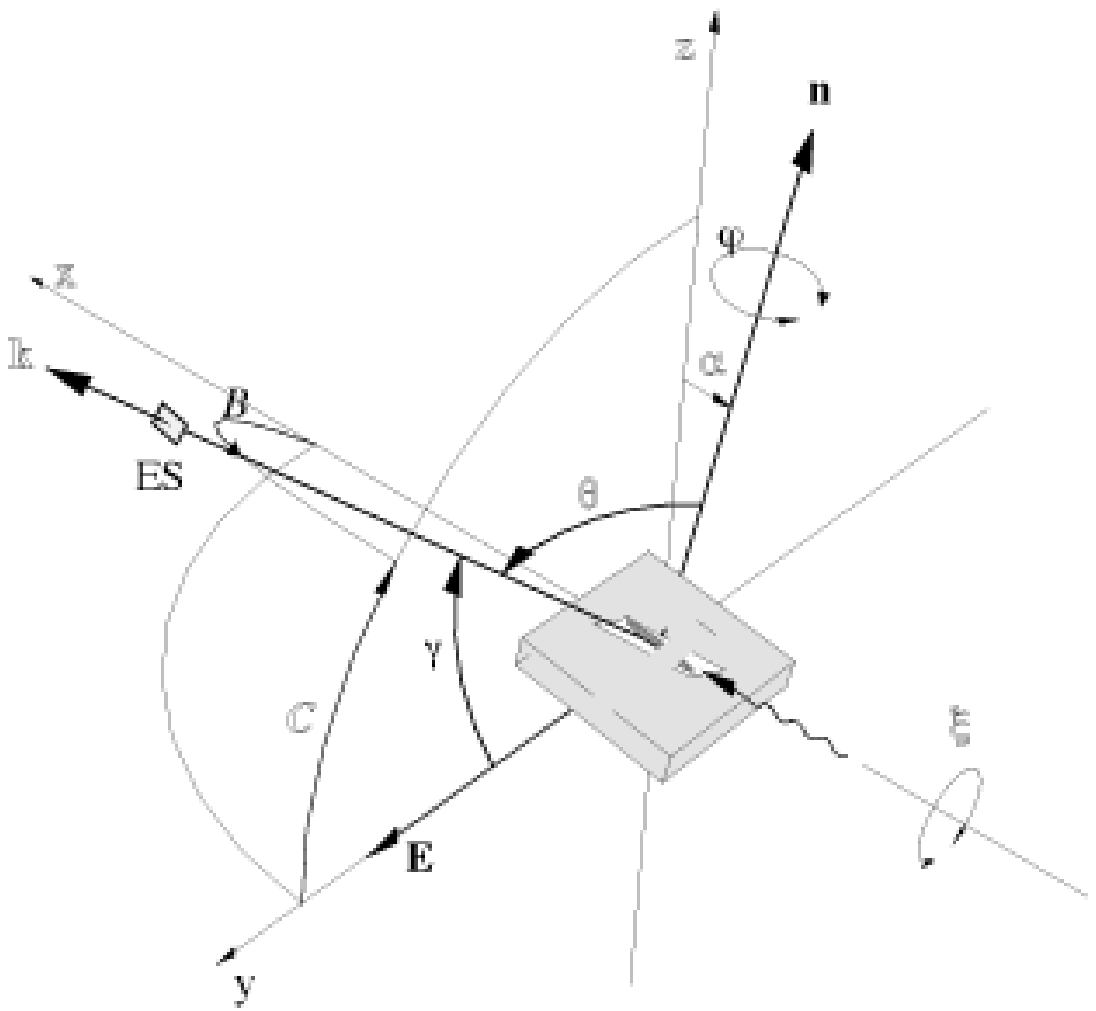, width=.45\textwidth}
\end{center}
\caption{Degrees of freedom in the choice of the scattering geometry 
for the ALOISA experimental station. 
The rotating electron 
analyzer is represented by its entrance slits $ES$; 
the projection of the 
slits onto the sample surface is drawn as a white parallelogram, 
defining the portion of surface contributing to the 
electron yield. The X-ray beam footprint on the surface (illuminated 
area) is shown as a shadowed ellipsis. The manipulator rotations 
$\alpha$, $\phi$ and $\xi$ are used to select independently the
grazing angle, the azimuthal orientation and the orientation of the 
surface with respect to the beam polarization vector $\vec{E}$. The 
analyzer rotations $B$ and $C$ allows one to survey most of the sky 
over the sample surface. The polar angle $\theta$ (emission angle 
referred to the surface normal $\hat{n}$) and the $\gamma$ angle 
(between the photoelectron wavevector $\vec{k}$ and $\vec{E}$) are 
also shown.} 
\label{geometry}
\end{figure}

A schematic picture of the angular degrees of freedom available in the 
determination of the scattering geometry is shown in Fig.\ref{geometry}. 
The manipulator rotation $\phi$ is used to select the azimuthal 
orientation of the surface, without altering the grazing angle 
which is selected by the rotation $\alpha$.
The surface normal can be rotated around the beam 
axis by $\pm180^{\circ}$  with the manipulator rotation $\xi$: at $\xi =0^{\circ}$ 
 the surface normal versor $\hat{n}$ lies in the plane normal to the electric 
field vector $\vec{E}$.

The rotation $B$ of the electron analyzer frame sets the deflection angle between 
the X-ray beam and the photoelectron wavevector $\vec{k}$, with a 
clearence of $\pm115^{\circ}$. Furthermore, the frame rotation axis 
$\hat{B}$ can 
be oriented in the $y-z$ plane of the laboratory reference system 
exploiting the $C$ rotation of the whole experimental chamber 
(independently of the manipulator settings). In particular, $\hat{B}$ is 
related to the rotation $C$ as
$\hat{B}=\hat{z}\cos C - \hat{y}\sin C$.

According to these definitions, the polar emission angle $\theta$ 
(defined by $\vec{k}$ and the surface normal direction $\hat{n}$) is 
defined as:

\begin{equation}
    \cos\theta(\xi,\alpha,B,C)= -\cos B \sin \alpha  +    
    \sin B \cos C \cos \alpha \sin \xi + 
    \sin B \sin C \cos \alpha \cos \xi,
\label{eq:polar}
\end{equation}

which simplifies to $\theta=90^{\circ}-(B-\alpha )$ if the analyzer is rotated 
in the plane containing the surface normal, which is often the case if 
the projection of the photoelectron wavevector onto the surface 
plane is to be mantained along a specific azimuthal direction, as in 
ED polar scans.

In the following we describe the factors contributing to the measured 
$ISO$ electron yield.

\paragraph{\textit{Atomic cross--section}}

In the dipole approximation (we will not consider multipolar 
expansion coefficients in the following applications), 
the atomic differential cross--section  
$d\sigma_{nl}/d\Omega$ for the $nl$ initial state of the 
photoemission process takes the well known
analytical expression,

\begin{equation}
    \frac{d\sigma_{nl}}{d\Omega}(\beta ; \gamma) \propto[1+\frac{1}{2}\beta (3\cos^{2}\gamma-1)],
\label{eq:cross}
\end{equation}

where $\beta(nl,h\nu)$ is the asimmetry parameter in 
the matrix 
elements for the photoemission process \cite{yeh} and 
$\gamma$ is the angle between the 
directions of the polarization vector $\vec{E}$ and the 
photoelectron wavevector $\vec{k}$ (see Fig.\ref{geometry}). 
The value of 
$\cos\gamma=\sin B \cos C$ is determined only  by the $B$ and $C$
rotations of the electron 
analyzer, independently from the sample surface orientation. 

\paragraph{\textit{Escape depth}}

The effect of inelastic scattering on the probability of escape of 
photoelectrons from the surface of condensed matter samples is 
described by means of the inelastic mean free path (IMFP), whose variation as 
a function of the electron kinetic energy is well known and can be 
evaluated by analytical formulas\cite{tanuma}.

The flux of electrons emitted at 
 a depth $z$ and detected at a polar angle $\theta$ from 
the surface normal will be reduced by inelastic scattering according 
to the Beer-Lambert relationship
$I(z,\theta)=e^{-\frac{z}{\lambda cos \theta}}$.

Integration over the depth from the surface yields the angular 
dependence expected for an emitting slab of matter of thickness $D$
\begin{equation}
I_{IMFP}(D,\lambda ;\theta)=\int_{0}^{D}I(z,\theta) \: dz = \lambda  \:
cos\theta \: [1-e^{-\frac{D}{\lambda \: cos\theta}}]
\label{eq:thickness}
\end{equation}
which corresponds to the well known $\sim cos(\theta)$ behaviour in the 
limit $D \rightarrow\infty$ of an homogeneous  semi-infinite emitting volume.

The same description can be used to describe the reduction of the 
photoemission intensity caused by a non--emitting layer, i.e. with 
different chemical composition, above the 
emitting region of the sample.
Considering an emitting layer of thickness $D$ below a non-emitting 
one of thickness $D^{\prime}$, the full formula can be written as:

\begin{equation}
I_{IMFP}(D,\lambda,D^{\prime},\lambda^{\prime} ;\theta) = \lambda  \:
cos\theta \: [1-e^{-\frac{D}{\lambda \: cos\theta}}] \: 
e^{-\frac{D^{\prime}}{\lambda^{\prime} \: cos\theta}}
\label{eq:doublethickness}
\end{equation}

Further, the limited penetration of 
the X-ray beam can be taken into account: this yields a very 
small correction, since the IMFP of 
the photoelectrons at the kinetic energies typical of photoemission 
experiments (50-1000~eV) is of the order of 10~\AA, while the characteristic 
penetration depth of the X-ray photons is normally larger than 50~\AA.

\paragraph{\textit{Surface roughness}}

The angular distribution of the photoelectron intensity, specially for 
high values of polar angle, is also affected by the 
surface roughness of the emitting layer. Models can be set up, to take 
into account the effects of shadowing of the photoelectron flux at 
grazing emission, where the details of the resulting formulae 
depend on the details of the surface morphology \cite{litt}.

A simple statistical model, which can be used in a phenomenological 
way, has been described in Ref.~\cite{Yako}.
The rough surface is described in terms of an isotropic stationary 
random function $z(x,y)$ of the in--plane surface coordinates with  a mean 
value $<z>=0$.  The distribution of heights is assumed to 
be normal, so that the probability 
density for $z$ is
$f(z)=\frac{1}{\sqrt{2\pi}\:\sigma}\: e^{-\frac{z^{2}}{2 \sigma^{2}}}$.

After a statistical analysis of the average shadowing by the surface 
protrusions, an analytical expression $I_{R}(\delta,\theta)$ for the dependence 
on the 
polar angle $\theta$ is derived. Details of the formulae can be found 
in Ref.\cite{Yako}. Qualitatively speaking,
$I_{R}$ assumes a  constant unit value near 
normal emission, and drops to zero for $\theta \rightarrow 90^{\circ}$,
the steepness of the decay being determined by the 
amount of surface roughness. 
The shadowing can be 
completely neglected up to a maximum 
take--off value of the polar angle $\arctan(\delta)$, so that the
phenomenological parameter $\delta$ can be used as an 
effective marker of the surface roughness. 

\paragraph{\textit{Surface illumination}}

Due to the high angular resolution of the ALOISA analyzers 
(acceptance angle of 1$^{\circ}$, FWHM), the angular smearing of the $ISO$ 
function originated by the angular acceptance can be neglected, 
although it must be taken into consideration in the simulation of the ED pattern $\chi$. 
On the other hand, the interplay between the size of the field of view 
and the illuminated portion area of the surface must be carefully taken into account. 
All the experiments at ALOISA are performed at 
grazing incidence, i.e. the 
value of the grazing angle $\alpha$ of the photon beam with respect to 
the sample surface is always in the $[0-10^{\circ}]$ range.  
Further, $\alpha$ is often set to values as low as a few degrees in 
order to exploit the increase of the signal from the surface region 
when total reflection condition are satisfied.
Even if 
the transverse width $\Gamma_{beam}$ of the focussed photon beam (FWHM) 
is narrow, the beam footprint on the surface is 
 elongated in the beam direction. 
The  width of the illuminated surface area $\Gamma(\alpha)=\Gamma_{beam}/ \sin 
\alpha$ can easily exceed the width of the surface area representing 
the projection of the analyzer slits ($1\times 4 
mm^{2} $), specially when 
working in very grazing conditions ($\alpha_{in}\sim 1^{\circ}$) and 
in transverse magnetic 
polarization, $\xi = 90^{\circ}$, where  $\Gamma_{beam}$ has its 
maximum value of $\sim150~\mu$m. This polarization setting is indeed the 
most frequently used, since it takes the nodal planes of
the atomic cross-section close to the surface horizon.

\begin{figure}[tbp]
\begin{center}
    \epsfig{file=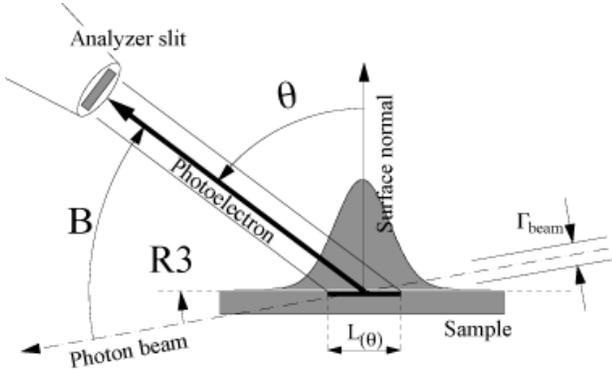, width=.45\textwidth}
\end{center}
\caption{ Schematic model of the intensity distribution of the 
photon beam illuminated area on the sample surface, for 
polar scans taken rotating the analyzer in the scattering plane. The 
projection of the analyzer slit on the surface 
$L(\theta)=\Delta/\cos\:\theta$ provides 
the integration limits in Eq.~\ref{eq:ill}.} 
\label{gauss}
\end{figure}

The beam footprint on the surface can be considered as an unidimensional
Gaussian  intensity distribution 
$exp\left\{ -\frac{x^{2}}{2 \Gamma^{2}(\alpha)} \right\} $ 
(see Fig.~\ref{gauss}), since the transverse distribution is always completely 
integrated by the analyzer slits. 

The integration of the photoemission intensity over the portion of illuminated 
area of the sample surface then results in the intensity factor 
\begin{equation}
    I_{ILL}(\Gamma_{beam}, \alpha; B, C)=\int_{-L/2}^{L/2}exp\left\{ -\frac{x^{2}}{2 
    (\Gamma (\alpha)/2.35)^{2}} \right\} dx
\label{eq:ill}
\end{equation}

The width $L$ of 
the portion of illuminated area, which overlaps to
 the projected area of the analyzer slits onto the sample 
surface, can be derived analytically, for any orientation of the sample 
surface and detector position. 
For a generic scattering configuration, $L$ is a non-trivial function 
of the experimental angles $B,C,\xi$ and $\alpha$, and can be also
limited by the finite sample size. The complete description of the 
general case is given in the Appendix. For simple polar scans,
in which the analyzer is rotated in the scattering plane defined by the 
photon wavevector and the surface normal , $L$ takes the minimum value
between the sample size and the projected slit width on the surface 
$\Delta/\cos\theta$. (see Fig.~\ref{gauss})

\paragraph{\textit{Complete $ISO$ formula}}

The complete formula combines all the described factors:

\begin{equation}
    ISO=A\cdot \frac{d\sigma_{nl}}{d\Omega}(\beta)\cdot 
    I_{IMFP}(D/\lambda, D^{\prime}/\lambda^{\prime})\cdot 
    I_{R}(\delta) \cdot
    I_{ILL}(\Gamma_{beam},\alpha),
\label{eq:ISO}
\end{equation}
where, for the sake of clarity, only the fitting parameters have been explicitely 
reported in the arguments of each correction factor.
The cross--section asymmetry parameter $\beta$ 
and the IMFP $\lambda$, $\lambda^{\prime}$ are set to calculated values. 
$A$ is a scale factor to be determined by fitting the data as 
well as the roughness parameter $\delta$. The X-ray beam width 
$\Gamma_{beam}$ and the grazing angle $\alpha$ are usually set to 
the nominal values in the first step of the iterative fitting 
procedure. The emitting layer thickness $D$ and the non-emitting overlayer 
thickness $D^{\prime}$
are also used as fitting parameters, unless an independent 
measurement is available for them (for instance by means of X-ray 
specular reflectivity\cite{bruno}).

\section{Data analysis}

\begin{figure}[tbp]
\begin{center}
    \epsfig{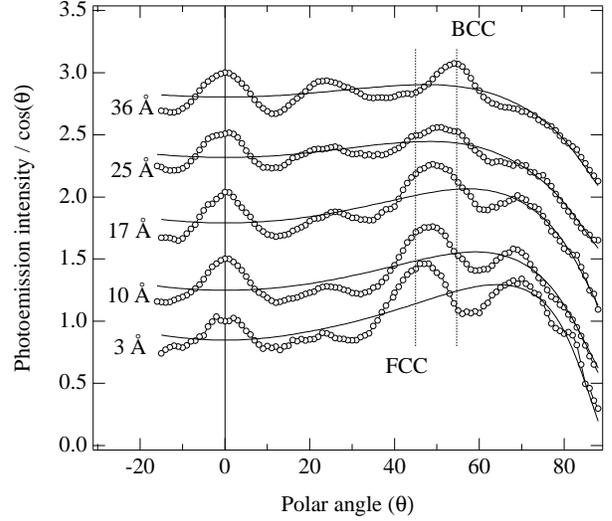}
\end{center}
\caption{ Polar scans taken for the Fe Auger LMM line from 
a growing crystalline Fe 
overlayer on Cu$_{3}$Au(001). The thickness dependence of the ISO 
distribution, also shown in the picture, has been highlighted by 
dividing both the data and the ISO curves by $\cos\:\theta$. The 
thicknesses have been independently determined by x-ray reflectivity.} 
\label{thickness}
\end{figure}

In Fig.~\ref{thickness}  a series of polar scans of the Fe Auger LMM line 
taken for a few Fe films  on Cu$_{3}$Au(001) are shown. 
For each film thickness, the experimental curve is compared with the 
calculated $ISO$ distribution.

In order to emphasize  the dependence on the emitting 
thickness $D$ (see Eq.~\ref{eq:thickness}), the curves were divided by 
a $\cos\theta$ factor.
Besides the structural transition from fcc(100) to 
bcc(100)$R45^{\circ}$~  
(witnessed by the angular shift of the main forward focusing 
directions), the weight of the thickness of the emitting layer on 
the $ISO$ shape is remarkable,
 particularly at the highest values of polar emission.
 The calculations were performed using 
the nominal Fe overlayer thicknesses (as determined by X-ray 
reflectivity) and an IMFP of $13.4$~\AA.
The roughness parameter $\delta$ was used as a free parameter and 
adjusted with a minimization procedure to the experimental data. 
For all the considered thicknesses,
 it fell in the range $\delta=1.0\pm 0.1$ (maximum 
un-shadowed takeoff angle $\sim 45^{\circ}$).

\begin{figure}[tbp]
\begin{center}
    \epsfig{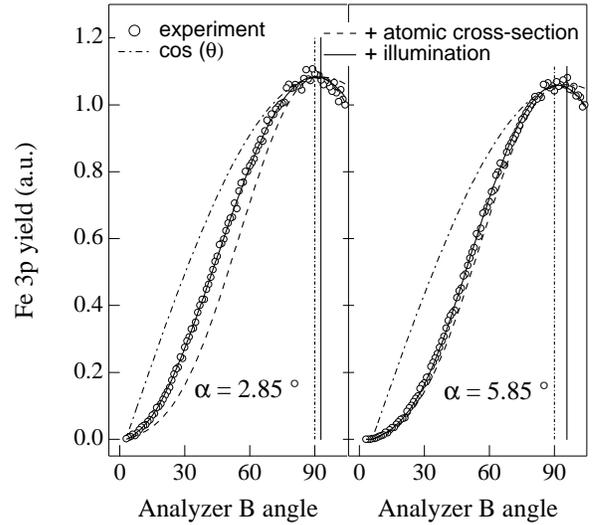}
\end{center}
\caption{Polar scans of Fe 3p intensity (at a kinetic energy of 803~eV)
from the surface of a policrystalline Fe sample 
are compared with calculated ISO functions, built by successive 
inclusion of the different factors (with $\beta$=1.485 for the atomic 
cross section factor and $\Gamma _{beam}$=150~$\mu$m for the 
illumination factor). The full ISO function fits the 
experimental data with the same set of parameters (see text) for both the scans, 
taken at different value of grazing angle $\alpha$. The vertical 
full line indicates the surface normal direction, while the 
vertical dot-dashed line is the direction of the photon beam electric field.} 
\label{illumination}
\end{figure}

To enlight the role of the illumination correction factor we show 
here data taken from the surface of a polycrystalline Fe sample, where the 
anisotropy $\chi$ component is absent and the ED pattern can be simply 
reproduced by the $ISO$ component. 
The polar scans of the Fe 3p photoelectron intensity (taken at a kinetic 
energy of 803~eV) 
are shown in Fig.~\ref{illumination} in comparison to a few calculated 
$ISO$ functions, built by successive 
inclusion of inelastic scattering, atomic cross--section and 
illumination correction factors.  
The complete $ISO$ component accurately fits the 
experimental data with the same set of parameters for both the scans. 
The illumination factor $I_{ILL}$ was calculated using 
$\Gamma_{beam}=150$~$\mu$m and $\alpha$ set to the nominal values, as 
indicated in the Figure. 
The asymmetry parameter in the cross-section factor was set to 
$\beta=1.485$, as calculated using one of the standard 
formulae \cite{tanuma}.
As expected, the weight of the illumination factor on the $ISO$ function 
 is more significant at lower grazing angles. 
Small values of the grazing angle $\alpha$ are usually chosen in the 
experiment settings in order to exploit the enhancement of the surface 
signal, obtained when $\alpha$ is as small as the critical 
value for total external reflection (of the order of a few degrees for 
soft X-rays). 

As far as the structural analysis is concerned, 
 the $\chi$ component is often extracted  by substracting an $ISO$ component, 
as obtained by interpolation of the 
ED pattern with a polynomial or a cosine function. The comparison 
between the so obtained experimental anysotropy and simulations 
based on structural models is then approached at a later stage.

\begin{figure}[tbp]
\begin{center}
    \epsfig{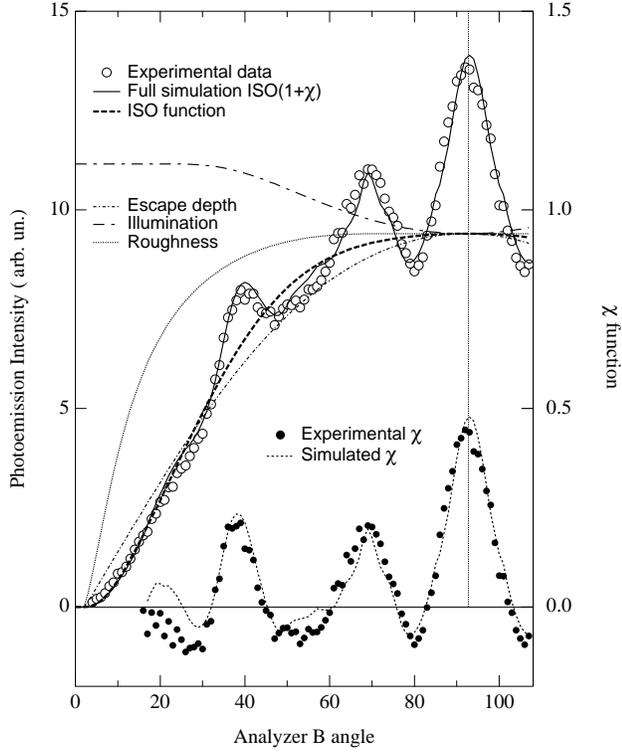}
\end{center}
\caption{Fe Auger LMM polar scan 
taken for the 36~\AA~ Fe film on Cu$_{3}$Au(001). The experimental ED 
pattern
is shown together with the full simulation $ED_{sim}$ and the correponding $ISO$ 
function. At the bottom, $\chi_{exp}=(ED_{exp}-ISO)/ISO$ and 
$\chi_{sim}=(ED_{sim}-ISO)/ISO$ are compared.
Three additional curves are added, as obtained recalculating the $ISO$ 
including only one factor at a time. The cross--section contribution 
is not shown, since we approximated the Auger emission with an 
isotropic distribution.
The normal emission direction is indicated by the vertical line.}
\label{defISO}
\end{figure}

Here, we propose the direct 
fitting of the complete ED experimental pattern, where the simulated 
ED curve is built up using both the multiple scattering ED calculation 
and the $ISO$ function, so that
$ ED_{sim}=ISO (1+\chi_{sim})$. 
In Fig.~\ref{defISO} we present an example of full fit to a polar scan 
taken on the Fe Auger LMM line,  
for a 36\AA~ Fe film  on Cu$_{3}$Au(001). The data are presented as a 
function of the analyzer $B$ angle (the normal emission direction, 
i.e. the zero for the polar angle $\theta$, is indicated by the vertical line).
The $\chi_{sim}$ component was calculated for a structural model 
by assuming the lateral lattice constant $a = 2.830$~\AA, 
as measured by in-plane 
X-ray diffraction \cite{bruno}, and the vertical parameter $c$ to be 
determined by the fitting procedure.
The $ED_{sim}$ was fitted to the experimental $ED_{exp}$, 
using in this case the roughness parameter $\delta$ and the X-ray beam width 
$\Gamma$ as fitting parameters. 
The layer thickness $D$ 
was set to the  value provided by the X-ray reflectivity calibration. 
In fact, being $D$ already as large as 
$3\lambda_{IMFP}$, it would have yielded a contribution very similar to the 
$D/lambda\rightarrow\infty$ limit, strongly reducing the reliability 
of its determination by the minimization of the ED fitting procedure.
An additional scale factor was also added to $\chi_{sim}$  
to take into account the ratio between the $\chi$ and $ISO$ amplitudes.

This procedure was iterated over a series of ED calculations, so that
for each structure--dependent
$\chi_{sim}$, the corresponding best fit $ISO$ parameters can be 
found, thus eliminating any 
arbitrary {\it a priori}  choice of the $ISO$ shape.
The best fit parameters of the $ISO$ component were found to be 
$\delta=0.9\pm 0.1$ and $\Gamma=160\pm20$~$\mu$m.

Three additional curves are added in the Figure, as obtained by 
calculating the $ISO$ component
including only one factor each time. 
  As anticipated before, the contribution due to the 
electrons escape depth is very close to the limit 
$I_{IMFP} \sim \cos\:\theta$. At low emission angles, the $ISO$ shape 
is strongly affected by the illumination and roughness factors.
The cross--section contribution 
is not shown, since the Auger emission was assumed to be 
isotropic. In fact,  an exact calculation would have required a detailed 
study of all the decay channels partecipating to the Auger  
process \cite{greber}, but the dependence on the angular momenta of 
the decay channels decreases as the kinetic energy increases, so that 
it becomes almost negligible at the present value of 698~eV 
\cite{idzerda}. 

At the bottom of Fig.\ref{defISO}, $\chi_{exp}=(ED_{exp}-ISO)/ISO$ and 
$\chi_{sim}=(ED_{sim}-ISO)/ISO$ are compared to highlight the 
quality of the fit. 
With the given lateral lattice spacing $a = 2.830$~\AA, we found the
ratio of the vertical to lateral spacing  to be $c/a = 1.03 \pm 
0.02$. The Fe film displays a structure very close to its bcc bulk one, 
$a = 2.86$~\AA~ and $c/a = 1$, the latter being 
recovered at a much higher thickness \cite{rochow}.
In general, this approach to the calculation of the $ISO$ component 
leads to a substantial improvement of the reliabilty of the structural 
model (with an indetermination not exceeding a few percents).

\section{Appendix}

The accurate description of the portion of illuminated area, which is 
included in the projected area of the analyzer slits onto the sample 
surface, is derived here for a generic scattering setup ($B,C,\xi,\alpha$).
For the sake of clarity, we will assume $\xi=0^{\circ}$, since only the 
mutual orientation of $\xi$ and $C$ is relevant.

First, the orientation of the slits in the laboratory reference system is defined 
by the vectors $\vec{B}_{slit}$ and $\vec{C}_{slit}$ which connect 
the center of the slit with the center of the short and the long 
sides of the slits, 
respectively. The $B$ and $C$ labels recall that the wide aperture is 
along the direction scanned by the $B$ angle, while the narrow one 
corresponds to a movement of the $C$ rotation.

\begin{eqnarray}
\vec{B}_{slit}=\frac{\Delta B}{2} \: \hat{k}\times 
\hat{B}_{axis} = \frac{\Delta B}{2} \: (\sin B, -\cos B \cos C, -\cos 
B \sin C) \\
\vec{C}_{slit}=\frac{\Delta C}{2} \: \hat{k}\times 
\hat{B}_{slit} = \frac{\Delta C}{2} \: (0, \sin C, \cos C),
\label{eq:Bp} 
\end{eqnarray}

where $\hat{k}=(\cos\:B, \sin\:B\:\cos\:C, \sin\:B\:\sin\:C)$ is the 
photoelectron waveversor, $\hat{B}_{axis}=(0, -\sin\:C, \cos\:C)$ is 
the orientation of the $B$ rotation axis and $\Delta B$ and 
$\Delta C$ are the full slit aperture in the two directions.

The displacement of the four vertices of the slit from the center of the 
slit in the laboratory reference system is then
\begin{equation}
    \vec{V}_{j}=\pm \vec{B}_{slit} \pm \vec{C}_{slit}; j=1,2,3,4.
\label{eq:V}
\end{equation}

The projection of the slit on the sample surface is in general a 
parallelogram, whose vertices are in the following positions in the 
laboratory reference system
\begin{equation}
    \vec{S}_{j}=-\frac{\hat{n}\cdot \vec{V}_{j}}{\hat{n}\cdot \hat{k}} 
    \hat{k} + \vec{V}_{j},
\label{eq:S}
\end{equation}

where $\hat{n}$ is the surface normal waversor.
The corresponding $\vec{S}^{\prime}$ positions in the (2-Dimensional) surface 
reference system are
\begin{eqnarray}
    \vec{S}_{j}^{\prime}=(\vec{S}_{j}\cdot \hat{x}_{tilt}, 
    \vec{S}_{j}\cdot \hat{y}_{tilt}),\\
    \hat{x}_{tilt}=(\cos\:\alpha, 0, \sin\:\alpha),  \\
    \hat{y}_{tilt}=\hat{y}_{Lab}. 
\label{eq:Sstar}    
\end{eqnarray}    

The width $L(B,C,\alpha)$ of the portion of illuminated area falling 
inside the parallelogram 
can be computed as the minimum among the
\begin{equation}
    L_{j}=2\left| 
    \frac{\vec{S}_{j+1,x}^{\prime}-\vec{S}_{j,x}^{\prime}}{\vec{S}_{j,y}^{\prime}
    -\vec{S}_{j+1,y}^{\prime}}\vec{S}_{j,y}^{\prime}  + 
    \vec{S}_{j,x}^{\prime}\right|; j=1,2,3,4;
\label{eq:zeri}
\end{equation}
which are the crossing points of the parallelogram sides with the 
$y=0$ plane on the surface (we recall that the X-ray spot on the sample 
can be considered as monodimensional). The integration 
lenght on the surface may be further limited by the surface physical length.

\newpage
\narrowtext
\twocolumn

\end{document}